\documentclass[aps,pra,english,showpacs,twocolumn]{revtex4}
\usepackage[english]{babel}
\usepackage[ansinew]{inputenc}
\usepackage{amsmath}
\usepackage{graphicx}

\newcommand{\kt}{\tilde{\kappa}}
\newcommand{\kh}{\hat{\kappa}}

\begin{document}

\title{Polarization squeezing by optical Faraday rotation}

\author{Jacob F. Sherson and Klaus M\o lmer}
\affiliation{Danish National Research Foundation Center for Quantum Optics\\Department of Physics and
Astronomy, University of Aarhus\\DK 8000 Aarhus C, Denmark}

\date{\today}

\begin{abstract}
  We show that it is possible to generate continuous-wave fields and
  pulses of polarization squeezed light by sending classical, linearly
  polarized laser light twice through an atomic sample which causes an
  optical Faraday rotation of the field polarization. We characterize
  the performance of the process, and we show that an appreciable
  degree of squeezing can be obtained under realistic physical
  assumptions.
\end{abstract}

  \maketitle

Highly squeezed states of light are valuable for numerous quantum
information protocols and high precision metrology . Several
techniques exist for the generation of squeezed states. In
\cite{UAndersenPolSq} the non-linear phase evolution due to the Kerr
effect in an optical fiber was used to produce 5.1 dB of polarization
squeezing. Cold atomic samples in high-finesse cavities cause similar
Kerr-like effects, and 1.5 dB of quadrature squeezing
\cite{Lambrecht96} and $\sim 0.5$ dB of polarization squeezing
\cite{giacobinoPRL} has been observed. The most well established
technique to date, however, is to use non-linear crystals in very good
cavities (e.g. $\sim 7$ dB in \cite{furusawaSq}).
In this Letter we show, however, that by reflecting a light beam so
that it interacts twice through the off-resonant Faraday-type
interaction with an atomic sample, a simple, robust, and efficient
source of strongly squeezed light is obtained, which may well
outperform the above mentioned schemes.

The optical Faraday rotation of light passing through a spin polarized
atomic medium has been applied in a number of recent experiments to
demonstrate entanglement and squeezing of atomic samples
\cite{julsgaardEntanglement,MabuchiSpinSqueezing}, atomic quantum
memories for light \cite{julsgaardQmem}, and teleportation of quantum
states between light and matter \cite{shersonTeleportation}. In these
experiments, the collective atomic population distribution on internal
states is a quantum degree of freedom, and the measurement of the
polarization of the light after the interaction alters the collective
atomic quantum state. Measurements play an important role in the
experiments reported in
\cite{julsgaardEntanglement,MabuchiSpinSqueezing,julsgaardQmem,
  shersonTeleportation}, but it has also been shown that multiple
interactions between a light pulse and an atomic medium
\cite{KlemensMultPass,shersonqubit,fiurasek,muschik} suffice to
effectively couple atoms and fields e.g. resulting in interspecies
entanglement, atomic squeezing, or transfer of quantum states.

Our proposed physical set-up is sketched in Figure 1. In part a) we
depict an atomic gas, spin polarized perpendicular to the plane of the
figure, through which a cw beam or a pulse of linearly polarized light
is transmitted twice from different directions. In part b) of the
figure we suggest an implementation with two oppositely polarized
samples which are both traversed twice by the laser beam.  As will
become clear below, the desired dynamics arises solely in the limit in
which the light field passes through the atoms from both directions
simultaneously.

\begin{figure}[t]
\includegraphics[width=0.40\textwidth]{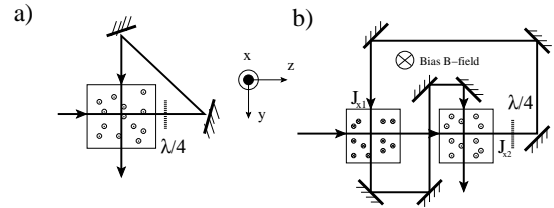}
\caption{\small
  Physical setup for production of squeezed light. In panel a), a
  coherent cw beam or pulse of linearly polarized light is transmitted
  through a spin polarized atomic gas, its polarization is rotated, and it is
  transmitted through the gas a second time from a different
  direction.The output field is polarization squeezed. In panel b),
  two oppositely polarized gases located in a constant B-field are
  traversed by the light field twice. The output field is polarization
  squeezed at the Larmor frequency sidebands.}
 \label{fig:setup}
\end{figure}

We treat the atomic ground state as an effective spin-1/2 system and
introduce the collective spin vector $\hat{\vec{J}}$. The
polarization degrees of freedom of the linearly polarized optical
field are accounted for by the Stokes vector$\hat{\vec{S}}$, and
both of these effective spin operators obey the usual angular
momentum commutator relations. As discussed in e.g.
\cite{julsgaardEntanglement,julsgaardQmem} the Faraday interaction
$\propto \hat{S}_z\hat{J}_z$ causes different phase shifts on the
$\sigma^+$ and $\sigma^-$ circularly polarized field components due
to the population difference on the atomic magnetic sub-states,
equivalent to a rotation of the linear optical polarization.
Conversely, the light induced AC-Stark shift imposes a phase
difference between the atomic magnetic sub-states equivalent to an
optical spin rotation around the z-axis proportional to the
intensity difference between $\sigma^+$ and $\sigma^-$ field
components.

For atoms which are spin polarized along the $x$-axis, it is useful
to apply the Holstein-Primakoff approximation
\cite{HolsteinPrimakoff}, where the macroscopic $x$-component of the
collective spin is treated as a c-number, and where the
dimensionless variables
$(x_{at},p_{at})=(\hat{J}_y,\hat{J}_z)/\sqrt{J_x}$ obey the
canonical commutator relation $[x_{at},p_{at}]=i$ (here, and
throughout, we set $\hbar=1$).  The minimum variances
$\textrm{Var}(x_{at})=\textrm{Var}(p_{at})=1/2$ reflect the binomial
distribution of population on the spin-y and -z eigenstates. We also
apply the Holstein-Primakoff approximation to the time dependent
Stokes vector components representing the field at a given distance
along the direction of propagation, so that the operators
$(x_{ph}(t),p_{ph}(t))=(\hat{S}_y(t),\hat{S}_z(t))/\sqrt{S_x(t)}$
with dimension $1/\sqrt{\textrm{time}}$, obey the canonical
commutator relation $[x_{ph}(t),p_{ph}(t')]=i\delta(t-t')$.

The Faraday interaction Hamiltonian can be rewritten in the new
variables, $H= \kappa
p_{at}p_{ph}$\cite{julsgaardQmem,shersonTeleportation,KlemensMultPass,shersonqubit},
and it is observed to be of quantum non-demolition (QND) type, i.e.,
it preserves the atomic variable $p_{at}$ and the field variable
$p_{ph}$. The coupling strength is $\kappa^2=\gamma\alpha$, with the
optical depth on resonance $\alpha=N_A\frac{\sigma}{A}$ and the
spontaneous emission rate
$\gamma=\Phi\frac{\sigma}{A}\frac{\Gamma^2}{\Delta^2}$. $N_A$ is the
number of atoms that interacts with the light, $\sigma$ is the
resonant light scattering cross section of a single atom, $A$ is the
cross section of the optical beam, $\Phi$ is the photon flux, $\Gamma$
is the natural linewidth of the atomic transition, and $\Delta$ is the
detuning from the optical transition. $\kappa$ has the dimension
$1/\sqrt{\textrm{time}}$, and $\kappa^2$ will be the natural scale for
the frequency dependence of our results. As illustrated in Fig.
\ref{fig:setup} we intend to pass the field around and propagate
simultaneously through the gas from a different direction effectively
realizing also the Hamiltonian, $H= \kappa x_{at}x_{ph}$. Our analysis
of the combined interaction becomes more involved and the full
dynamics is no longer QND.

We shall apply an input-output formalism, where the field variables
$x_i(t),p_i(t)$ after the first exit of the atomic sample are given in
terms of the entrance variables $x_{in}(t),p_{in}(t)$ and the atomic
variables,
\begin{eqnarray}\label{pass1}
x_i(t)=x_{in}(t)+\kappa p_{at}(t)\nonumber \\
p_i(t)=p_{in}(t).
\end{eqnarray}

In (\ref{pass1}) we assume that the passage of light through the gas
is instantaneous, and hence the atomic and field variables can all be
evaluated at the same time. The light propagates and is reflected by
mirrors and is therefore subject to losses. These can effectively be
modelled by a transmission coefficient $\tau$, which is less than
unity, and noise terms $\rho F_{x(p)}(t)$, where $\rho^2=1-\tau^2$,
$[F_x(t),F_p(t')]=i\delta(t-t')$ and
$[F_x(t),F_x(t')]=[F_p(t),F_p(t')]=0$. We thus modify (\ref{pass1}) to
represent the fields prior to the second passage of the gas,
\begin{eqnarray} \label{damp}
x_i'(t)=\tau(x_{in}(t)+\kappa p_{at}(t))+\rho F_x(t) \nonumber \\
p_i'(t)=\tau p_{in}(t)+\rho F_p(t).
\end{eqnarray}
In this set of equations, we have ignored the time of propagation
between the two passes. This approximation amounts to the assumption
that the atomic state changes only little on the time scale of the
optical propagation, which will always be the case with realistic
physical parameters.

Before the second interaction between the field and the atoms the
field polarization is rotated $90^\circ$ (using a
$\lambda/4$-plate), and the atomic sample is approached from a
different direction, altogether realizing the Hamiltonian $H=\tau\kappa
x_{at} x_{ph}$,  where the $\tau$ factor accounts for the reduced
interaction strength because of the photon loss ($\Phi \rightarrow
\tau^2\Phi$) between the passages. We now express the output fields
$x_{out}(t), p_{out}(t)$ in terms of the atomic variables and the
intermittent field components (\ref{damp}):
\begin{eqnarray} \label{pass2}
x_{out}(t)=x_i'(t)\nonumber \\
p_{out}(t)=p_i'(t) -\kappa \tau x_{at}(t),
\end{eqnarray}
For mathematical convenience we have applied another $90^\circ$
rotation of the light after the second passage.

The atoms undergo dissipation, both due to the weak excitation by
the optical fields, which causes a small spontaneous emission rate
$\gamma$, and because we may implement optical pumping with a rate
$\gamma_p$ to retain the atomic macroscopic polarization along the
x-axis. The mean spin will thus obey $\frac{d J_x}{dt}=-\gamma
J_x+\gamma_p(N/2-J_x)$ with a macroscopic steady state value
$J_x=\frac{\gamma_p}{\gamma+\gamma_p}N/2$. The atomic variables
$x_{at}$ and $p_{at}$ decay with the rate
$\gamma_1=\gamma+\gamma_p$. The decay is accompanied by noise terms
$\sqrt{2\gamma_2} G_{x(p)}(t)$, where $G_{x(p)}(t)$ have the same
commutator properties as the $F_{x(p)}(t)$ operators introduced for
the field losses. Due to the division by $\sqrt{J_x}$ in the
Holstein-Primakoff approximation, we obtain the value
$\gamma_2=\gamma_1/(\frac{\gamma_p}{\gamma+\gamma_p}) =
(\gamma+\gamma_p)^2/\gamma_p$ for the strength of the noise term. If
there are other decoherence mechanisms present, $\gamma_1$ and
$\gamma_2$ of course have to be modified correspondingly.

The atomic interaction with the field at the two passages and the damping are described by the
equations
\begin{eqnarray}\label{atomic}
\frac{d}{dt}x_{at}(t)=\kappa p_{in}(t) -\gamma_1 x_{at}(t) +
\sqrt{2\gamma_2} G_x(t)\nonumber \\
\frac{d}{dt}p_{at}(t)=-\kappa \tau x_i'(t) - \gamma_1 p_{at}(t) +
\sqrt{2\gamma_2} G_p(t),
\end{eqnarray}
which have to be solved together with the field equations
(\ref{damp},\ref{pass2}).

To solve the equations under cw operation, we change to the frequency
domain, letting $h(t)=\frac{1}{\sqrt{2\pi}}\int e^{i\omega t}h(\omega)
d\omega$ for all operators $h=x_{at}, p_{at}, x_{in}, p_{in}, x_i,
p_i, x_{out}, p_{out},F_x, F_p, G_x, G_p$. This transformation changes
the argument from time to frequency in the field equations
(\ref{damp},\ref{pass2}), and it changes the differential equations
(\ref{atomic}) into the algebraic equations
\begin{eqnarray}\label{atomicfreq}
i\omega x_{at}(\omega)=\kappa p_{in}(\omega) -\gamma_1
x_{at}(\omega) +
\sqrt{2\gamma_2} G_x(\omega)\nonumber \\
i \omega p_{at}(\omega)=-\kappa \tau x_i'(\omega) - \gamma_1
p_{at}(\omega) + \sqrt{2\gamma_2} G_p(\omega).
\end{eqnarray}

We can now systematically express both the atomic variables and the
output field variables in terms of the input field and noise
operators, and we find by elementary operations,
\begin{equation} \label{xout}
x_{out}(\omega)=\frac{ (\gamma_1+i\omega)(\tau x_{in}+\rho
F_x(\omega))+\tau\sqrt{2\gamma_2}\kappa G_p(\omega)}
{\gamma_1+i\omega+\tau^2\kappa^2}.
\end{equation}
If we assume the vacuum values for the noise power spectrum of the
incident classical field and the noise variables $\langle
h(t)h(t')\rangle = \frac{1}{2}\delta(t-t') \Rightarrow \langle
h(\omega) h(\omega') = \frac{1}{2}\delta(\omega+\omega')$, the noise
spectrum of squeezing of the field is given by the expectation value
$\langle x_{out}(\omega)
x_{out}(\omega')\rangle=V_x(\omega)\delta(\omega+\omega')$, and we
obtain directly from (\ref{xout}),
\begin{equation} \label{noise}
V_x(\omega) = \frac{1}{2}\left( 1- \frac{2(\gamma_1-\gamma_2)\kappa^2\tau^2+\kappa^4\tau^4}
{(\gamma_1+\kappa^2\tau^2)^2+\omega^2}\right).
\end{equation}
The squeezing spectrum is a Lorentzian, with a width given by the
atomic decoherence rate, the coupling strength to the atoms and the
transmission efficiency of light between the two passages of the gas.
Interestingly, the effect of losses between the two light passages of
the gas only has the effect to modify the value of the interaction
strength, $\kt=\kappa\tau$.
\begin{figure}[t]
  \includegraphics[width=0.40\textwidth]{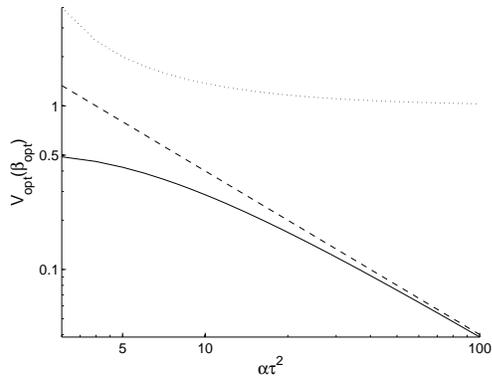}
\caption{\small
  The optimal degree of squeezing $V_\mathrm{opt}$, Eq.
  (\ref{eq:maxsqopt}) (solid line) approached asymptotically by the
  expression $4/(\alpha\tau^2)$ (dashed line) as functions of
  $\alpha\tau^2$.  Both curves assume the optimal strength of the
  optical pumping, $\beta=\gamma_p/\gamma$, which is shown as the
  upper dottted line.  }
 \label{fig:spec}
\end{figure}

Assuming that the dominant atomic decoherence mechanism is light
absorption, and writing $\kappa^2=\gamma\alpha$ and
$\gamma_p=\beta\gamma$, the maximum squeezing (at $\omega=0$) is:
\begin{equation}
  \label{eq:maxsq}
  V_x(0)=\frac{1}{2}\frac{(1+\beta)^2(\beta+2\alpha\tau^2)} {\beta(1+\beta+\alpha\tau^2)^2},
\end{equation}
Optimizing Eq. (\ref{eq:maxsq}), we get
\begin{equation}
  \label{eq:maxsqopt}
  V_\mathrm{opt}=\frac{1}{2}
\frac{(2\alpha\tau^2-1)^3}{(\alpha\tau^2-1)(1+\alpha\tau^2)^3}
\to\frac{4}{\alpha\tau^2}\qquad \alpha\tau^2\gg1
\end{equation}
for $\beta = \frac{1+\alpha\tau^2}{\alpha\tau^2-2}$. In Fig.
\ref{fig:spec} we plot Eq.  (\ref{eq:maxsqopt}) along with the optimal
value for $\beta$. As can be seen, degrees of squeezing of the order
of 10 dB should realistically be achieved for optical depths
$\alpha\approx 100$ which is routinely attained in cold atomic samples
as in \cite{MabuchiSpinSqueezing}.  In a BEC optical densities of up
to $10^3$ can be achieved potentially paving the way for unprecedented
degrees of squeezing. The high degrees of squeezing are attained while
the mean spin $J_x= \beta/(1+\beta)\cdot N/2$ is reduced by a factor
on order unity compared to its maximal value. For
$\alpha\tau^2\gg\beta\gg1$ the mean spin is only reduced slightly and
the non-optimal degree of squeezing approaches $\beta/(\alpha\tau^2)$,
which can still be significant.

The atoms can also be optically pumped into the macroscopically
polarized state and interact only with a finite duration pulse of
light, so that the reduction in the mean spin, with rate $\gamma$, is
small.  The equations (\ref{damp}-\ref{atomic}), with
$\gamma_1=\gamma_2=\gamma$ can be solved directly in the time domain,
expressing the time dependent output fields as integrals over the
incident field and noise operators. At this point, there are rich
possibilities to vary the incident field envelope and the mode
function identified with the output field with the purpose to optimize
the squeezing. For simplicity, we shall here assume an incident square
pulse, and we shall consider the amount of squeezing in a mode defined
by the same pulse envelope, i.e., we consider the single mode field
operators $q_T=\frac{1}{\sqrt{T}}\int_0^T q_{out}(t)dt$, where $q=x$
or $p$, and where $T$ is the duration of the light pulse. From our
full time dependent solution we obtain the output variances:
\begin{figure}[t]
\includegraphics[width=0.40\textwidth]{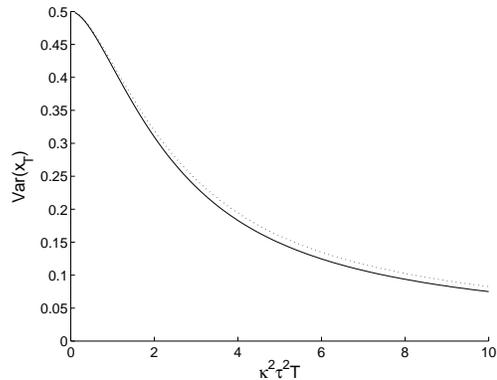}
\caption{\small
  The variance $\mathrm{Var}(x_T)$ after passage of a square pulse
  through the single cell set-up (see Fig. \ref{fig:setup} a)) The
  variance is shown as a function of $\kappa^2\tau^2 T$ 
  with finite damping $\gamma T=0.1$ (dotted line) and with no damping
  (solid line).}
 \label{fig:pulseXl}
\end{figure}
\begin{equation}
  \label{eq:varPlPulse}
  \mathrm{Var}(q_T)=\frac{1}{2}\left(1\mp
\frac{\kappa^4 \tau_q^2}{2\gamma_q^3 T}
\left(2\gamma_q T +1-(2 -  e^{-\gamma_q T})^2\right)\right)~,
\end{equation}
where $q=x,p$, $\tau_x=\tau^2$, $\tau_p=\tau$,
$\gamma_x=\gamma+\kappa^2\tau^2$, $\gamma_p=\gamma$, and where the
-(+) sign apply to the $x_T$($p_T$) component.  In Fig.
\ref{fig:pulseXl} we plot the attainable degree of squeezing as a
function of the dimensionless quantity $\kappa^2\tau^2T$. The
assumption of this calculation is that the mean spin is preserved, and
we show in the figure the results for finite damping $\gamma T=0.1$
and for no damping $\gamma T=0$, which provide quite equivalent
results. The attainable degree of sqeezing is clearly sizable. In the
limit of no damping, Eq.(\ref{eq:varPlPulse}) approaches the simple
expressions
 \begin{equation}
   \label{eq:pulseVar}
   \mathrm{Var}(x_T)= \frac{3+e^{-2\kh^2}-4e^{-\kh^2}}
{4\kh^2}~,~~
 \mathrm{Var}(p_T)=\frac{1}{2}+\frac{\kh^4}{6},
 \end{equation}
 where $\kh^2=\kappa^2T$. We note that the product of the two variance
 is 1/4 for $T=0$ but the squeezing of the $x$ component is
 accompanied by a growth of the uncertainty product for $\kappa^2
 T\gg1$.

\begin{figure}[t]
\includegraphics[width=0.40\textwidth]{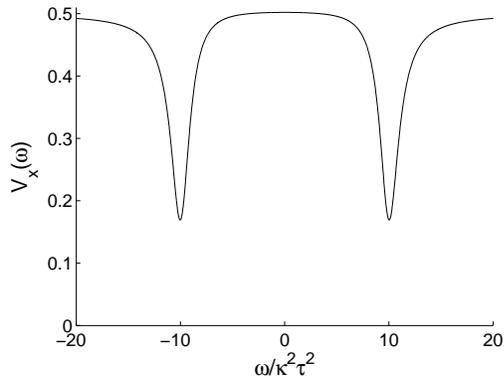}
\caption{\small
  Squeezing spectrum for continuous wave field transmitted twice
  through two atomic ensembles which are larmnor precessing at
  freqency $\Omega$. The parameters in the figure are
  $\Omega=10\kt^2$, $\gamma_1=0.1\kt^2$, and $\gamma_2=0.2\kt^2$
  (corresponding to $\gamma=\gamma_p$).}
 \label{fig:spec2cell}
\end{figure}

Our proposal can also be implemented with two oppositely oriented
atomic samples and a homogeneous magnetic field as illustrated in Fig.
\ref{fig:setup} b). To model this we introduce atomic variables
$p_{at,i}= J_{z,i}/\sqrt{J_x}$, and $x_{at,i}=\pm J_{y,i}/\sqrt{J_x}$,
where $i=1,2$. This gives equations of motion very similar to
equations (\ref{damp}-\ref{atomic}) and the Larmor precession caused
by the constant magnetic field is modelled by adding terms $\pm \Omega
x_{at,i}$ and $\mp \Omega p_{at,i}$ to the differential equations for
$p_{at,i}$ and $x_{at,i}$ respectively. Note that the rotation is in
opposite directions for the two atomic samples.
Solving the coupled equations we get {\small
\begin{equation}
  \label{eq:twocellX}
V_x(\omega)=\frac{1}{2}\left(1-\frac{4\kt^4(\gamma_1^2+\omega^2)+
4\kt^2(\gamma_1-\gamma_2)(\gamma_1^2+\omega^2+\Omega^2)} {\gamma_1^2
A^2+2A\gamma_1(\omega^2+\Omega^2)+(\omega^2-\Omega^2)^2 +
4\kt^4\omega^2}\right)~,
\end{equation}
} where $\kt=\kappa\tau$ and $A=\gamma_1+2\kt^2$. Again losses
between the two passages merely modify the interaction strength by a
factor as was also the case in the single cell implementation (Eq.
(\ref{noise})). In the limit of fast rotations
($\Omega\gg\gamma,\kt$) we get two well separated peaks in the
squeezing spectrum centered around $\omega=\pm\Omega$ as shown in
Fig. \ref{fig:spec2cell}.
Introducing $\omega=\pm \Omega+\delta$ and letting $|\Omega|\gg
\delta,\gamma_1$, Eq. (\ref{eq:twocellX}) reduces to a Lorentzian
dependence on the frequency off-set with respect to $\pm \Omega$
with the same parameters as Eq. (\ref{noise}) for a single sample
without rotations. It has previously been shown that the Faraday-QND
interaction of a single sample without a contant bias magnetic field
is regained for two oppositely oriented samples in the presence of a
bias field \cite{julsgaardEntanglement,julsgaardQmem}. Our result
shows that this correspondence extends to our non-QND situation,
when the dynamics is fast compared to the magnetic precession
frequency.

In summary, we have shown that both light pulses and cw light fields
passing twice through an atomic gas or through two atomic samples
become significantly squeezed. The width of the sqeezing spectrum is
governed by the coupling strength $\kappa^2$, and the optimum of
squeezing is controlled by the resonant optical depth. Realistic
estimates for these parameters in a number of current experiments
suggest that the attainable squeezing competes well with the
achievements of other schemes. Optical transmission losses after the
last atomic interaction can readily be modelled with a transmission
coeffecient and a noise term as in Eq.(\ref{damp}), and the relevant
single mode or continuous wave variances $V$ are modified according
to, $V \rightarrow \tilde{\tau}^2V+\tilde{\rho}^2\frac{1}{2}$.

Let us conclude with a brief discussion of the "simultaneous passage"
mechanism behind the squeezing.  By inserting Eq. (\ref{damp}) into
Eq.  (\ref{atomic}) it is clear that both during our cw and our longer
pulse transmission, $p_{at}$ is fed back onto itself with a negative
factor, which leads to a damping of the corresponding variance. In Eq.
(\ref{pass1}), $p_{at}$ is, in turn, mapped onto $x_i$ and hence onto
$x_{out}$, c.f. Eqs.(\ref{damp},\ref{pass2}), which gives rise to the
squeezed output state of light. For pulses short enough to allow a
complete passage in one direction before the passage in the other one,
it is the initial unsqueezed $p_{at}$ which maps onto the field
variable in the first interaction, and $x_{out}$ will thus be even
more noisy than in the input state. The simultaneous interaction
assumes that the travel time for the light between passes should be
much smaller than the pulse length or the reciprocal of the cw
squeezing bandwidth, which is readily fulfilled in experiments.

\bibliographystyle{apsrev}
\bibliography{squeezing}

\end{document}